\documentclass[%
 reprint,
 amsmath,amssymb,
 aps,
prb
]{revtex4-1}

\usepackage{graphicx}% Include figure files
\usepackage{dcolumn}% Align table columns on decimal point
\usepackage{bm}% bold math
\usepackage{upgreek} % non-italic greek letters in math mode

\begin{document}

%\preprint{APS/123-QED}

\title{Spin-Hall Magnetoresistance in Platinum on Yttrium Iron Garnet: Dependence on platinum thickness and in-plane/out-of-plane magnetization}

%--------------------------------Author------------------------------------------------------------------------------------------------------------------
%-------------------------------------------------------------Author-------------------------------------------------------------------------------------
\author{N. Vlietstra}
% \email{n.vlietstra@rug.nl}
\affiliation{ 
University of Groningen, Physics of nanodevices, Zernike Institute for Advanced Materials, Nijenborgh 4, 9747 AG Groningen, The Netherlands.%\\This line break forced with \textbackslash\textbackslash
}%

\author{J. Shan}
\affiliation{ 
University of Groningen, Physics of nanodevices, Zernike Institute for Advanced Materials, Nijenborgh 4, 9747 AG Groningen, The Netherlands.%\\This line break forced with \textbackslash\textbackslash
}%

\author{V. Castel}
\affiliation{ 
University of Groningen, Physics of nanodevices, Zernike Institute for Advanced Materials, Nijenborgh 4, 9747 AG Groningen, The Netherlands.%\\This line break forced with \textbackslash\textbackslash
}%

\author{J. Ben Youssef}%
\affiliation{ 
Universit\'e de Bretagne Occidentale, Laboratoire de Magn\'etisme de Bretagne CNRS, 6 Avenue Le Gorgeu, 29285 Brest, France.%\\This line break forced with \textbackslash\textbackslash
}%

\author{B. J. van Wees}
\affiliation{ 
University of Groningen, Physics of nanodevices, Zernike Institute for Advanced Materials, Nijenborgh 4, 9747 AG Groningen, The Netherlands.%\\This line break forced with \textbackslash\textbackslash
}%

\date{\today}% It is always \today, today,
             % but any date may be explicitly specified

\begin{abstract}

The occurrence of Spin-Hall Magnetoresistance (SMR) in platinum (Pt) on top of yttrium iron garnet (YIG) has been investigated, for both in-plane and out-of-plane applied magnetic fields and for different Pt thicknesses [3, 4, 8 and 35nm]. Our experiments show that the SMR signal directly depends on the in-plane and out-of-plane magnetization directions of the YIG. This confirms the theoretical description, where the SMR occurs due to the interplay of spin-orbit interaction in the Pt and spin-mixing at the YIG/Pt interface. Additionally, the sensitivity of the SMR and spin pumping signals on the YIG/Pt interface conditions is shown by comparing two different deposition techniques (e-beam evaporation and dc sputtering). 

\begin{description}

\item[PACS numbers]
72.25.Ba, 72.25.Mk, 75.47.-m, 75.76.+j
\end{description}
\end{abstract}


\maketitle

%------------------------------------------------------------------------------------------------------------------------------------------------------------------
%------------------------------------------------------------------------------------------------------------------------------------------------------------------
%------------------------------------------------------------------------------------------------------------------------------------------------------------------
%------Introduction------
\section{Introduction}
Platinum (Pt) is a suitable material to be used as a spin-current to charge-current converter due to its strong spin-orbit coupling.\citep{AndoIEEE2010} A spin current injected into a Pt film will generate a transverse charge current by the Inverse Spin-Hall Effect (ISHE), which can then be electrically detected. The ISHE has been used to detect for example spin pumping into Pt from various materials such as permalloy\citep{SaitohAPL2006} (Py) and yttrium iron garnet (YIG).\citep{ISHESaitohJAP,Kurebayashi2011nmat,CastelPRB} For the opposite effect, to use Pt as a spin current injector, a charge current is sent through the Pt, creating a transverse spin accumulation by the Spin-Hall Effect (SHE).\citep{SHESaitohPRL,SHEBuhrmanPRL,AzevedoAPL2011} %For most experiments where Pt is used to create or detect spin currents, possible magnetization of the Pt layer is neglected\citep{GoennenweinPtMagnetic}. Only a few groups investigate magnetic behaviour in the Pt layer, which might interfere with the ISHE and SHE experiments\citep{ChienPRL2012,BauerSMR}. 

\begin{figure}[b]
\includegraphics[width=8.5cm]{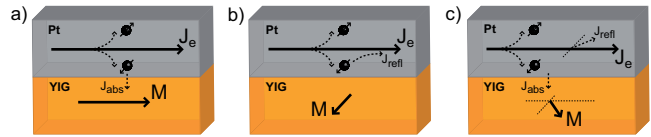}% Here is how to import EPS art
\caption{\label{fig:Fig1} 
Schematic drawing explaining the SMR in a YIG/Pt system. (a) When the magnetization $\vec{M}$ of YIG is perpendicular to the spin polarization $\vec{\sigma}$ of the spin accumulation created in the Pt by the SHE, the spin accumulation will be absorbed ($\vec{J}_{abs}$) by the localized moments in the YIG. (b) For $\vec{M}$ parallel to $\vec{\sigma}$, the spin accumulation cannot be absorbed, which results in a reflected spin current back into the Pt, where an additional charge current $\vec{J}_{refl}$ will be created by the ISHE. (c) For $\vec{M}$ in any other direction, the component of $\vec{\sigma}$ perpendicular to $\vec{M}$ will be absorbed and the component parallel to $\vec{M}$ will be reflected, resulting in a current $\vec{J}_{refl}$ which is not collinear with the initially applied current $\vec{J_e}$.
}
\end{figure}

Recently, Weiler et al.\citep{WeilerPRL2012} and Huang et al.\citep{ChienPRL2012} observed magnetoresistance (MR) effects in Pt on YIG and related those effects to magnetic proximity. These MR effects have been further investigated by Nakayama et al.\citep{BauerSMR} and they found and explained a new magnetoresistance, called Spin-Hall Magnetoresistance (SMR).\citep{BauerSMR,BauerTheorySMR}
% Explain SMR
A change in resistance due to SMR can be explained by a combination of the Spin-Hall Effect (SHE) and the Inverse Spin-Hall Effect (ISHE), acting simultaneously. When a charge current $\vec{J_e}$ is sent through a Pt strip, a transverse spin current $\vec{J_s}$ is generated by the SHE following $\vec{J_e} \propto \vec{\sigma}\times\vec{J_s}$,\citep{SHE,NiFePtdep,AzevedoPRB2011,Kajiwara2010nature} where $\vec{\sigma}$ is the polarization direction of the spin current. Part of this created spin current is directed towards the YIG/Pt interface. At this interface the electrons in the Pt will interact with the localized moments in the YIG as is shown in Fig. \ref{fig:Fig1}. Depending on the magnetization direction of the YIG, electron spins will be absorbed ($\vec{M} \perp \vec{\sigma}$) or reflected ($\vec{M} \parallel \vec{\sigma}$). 
By changing the direction of the magnetization of the YIG, the polarization direction of the reflected spins, and thus the direction of the additional created charge current, can be controlled. 
A charge current with a component in the direction perpendicular to $\vec{J_e}$ can also be created, which generates a transverse voltage. 

In this paper, the angular dependence of the SMR in Pt on YIG is investigated for different Pt thicknesses (3, 4, 8 and 35nm) and different deposition techniques (e-beam evaporation and dc sputtering), for applied in-plane as well as out-of-plane magnetic field sweeps, revealing the full magnetization behaviour of the YIG.\footnote{Nakayama et al.\citep{BauerSMR} also investigated the out-of-plane behavior of the SMR, but only for saturated magnetization directions, which are fully aligned to the applied field} All measurements are performed at room temperature. The magnitude of the SMR is shown to be dependent on the magnetization direction of the YIG, as well as on the Pt thickness, indicating its relation to the spin diffusion length. Also the used deposition technique is found to be an important factor for the magnitude of the measured signals.
 
%------------------------------------------------------------------------------------------------------------------------------------------------------------------
%------------------------------------------------------------------------------------------------------------------------------------------------------------------
%------------------------------------------------------------------------------------------------------------------------------------------------------------------
%------Experimental details------

% Sample description
\section{Sample characteristics}
\begin{figure}[h]
\includegraphics[width=8.5cm]{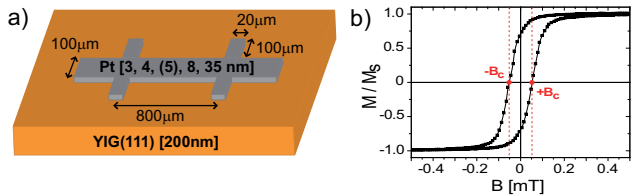}% Here is how to import EPS art
\caption{\label{fig:Fig2} 
(a) Schematics of the used Pt Hall bar geometry. (b) In-plane magnetic field dependence of the magnetization $M$ of the pure single-crystal of YIG. $B_c$ indicates the coercive field of 0.06mT.
}
\end{figure}

Pt Hall bars with thicknesses of 3, 4, 8, and 35nm were deposited on YIG by dc sputtering. Similar Pt Hall bars were also deposited on a Si/SiO$_2$ substrate, as a reference. Finally a sample was fabricated where a layer of Pt (5nm) was deposited on YIG by e-beam evaporation. Fig. \ref{fig:Fig2}(a) shows the dimensions of the Hall bars. The thickness of the deposited Pt layers was measured by atomic force microscopy with an accuracy of $\pm$0.5nm. The used YIG (single-crystal) has a thickness of 200nm and is grown by liquid phase epitaxy on a (111) Gd$_3$Ga$_5$O$_{12}$ (GGG) substrate. By using a vibrating sample magnetometer, the magnetic field dependence of the magnetization was determined, as shown in Fig. \ref{fig:Fig2}(b). The magnetic field dependence shows the same magnetization behaviour for all in-plane directions, indicating isotropic behaviour of the magnetization in the film plane, with a low coercive field of only 0.06mT. To saturate the magnetization of this YIG sample in the out-of-plane direction, an external magnetic field higher than the saturation field ($\mu_0M_s=0.176$T)\citep{CastelPRB} has to be applied.

%------------------------------------------------------------------------------------------------------------------------------------------------------------------
%------------------------------------------------------------------------------------------------------------------------------------------------------------------
%------------------------------------------------------------------------------------------------------------------------------------------------------------------
%------Results and discussion------
\section{Results and Discussion}
\subsection{In-plane magnetic field dependence}
\begin{figure}
\includegraphics[width=8.5cm]{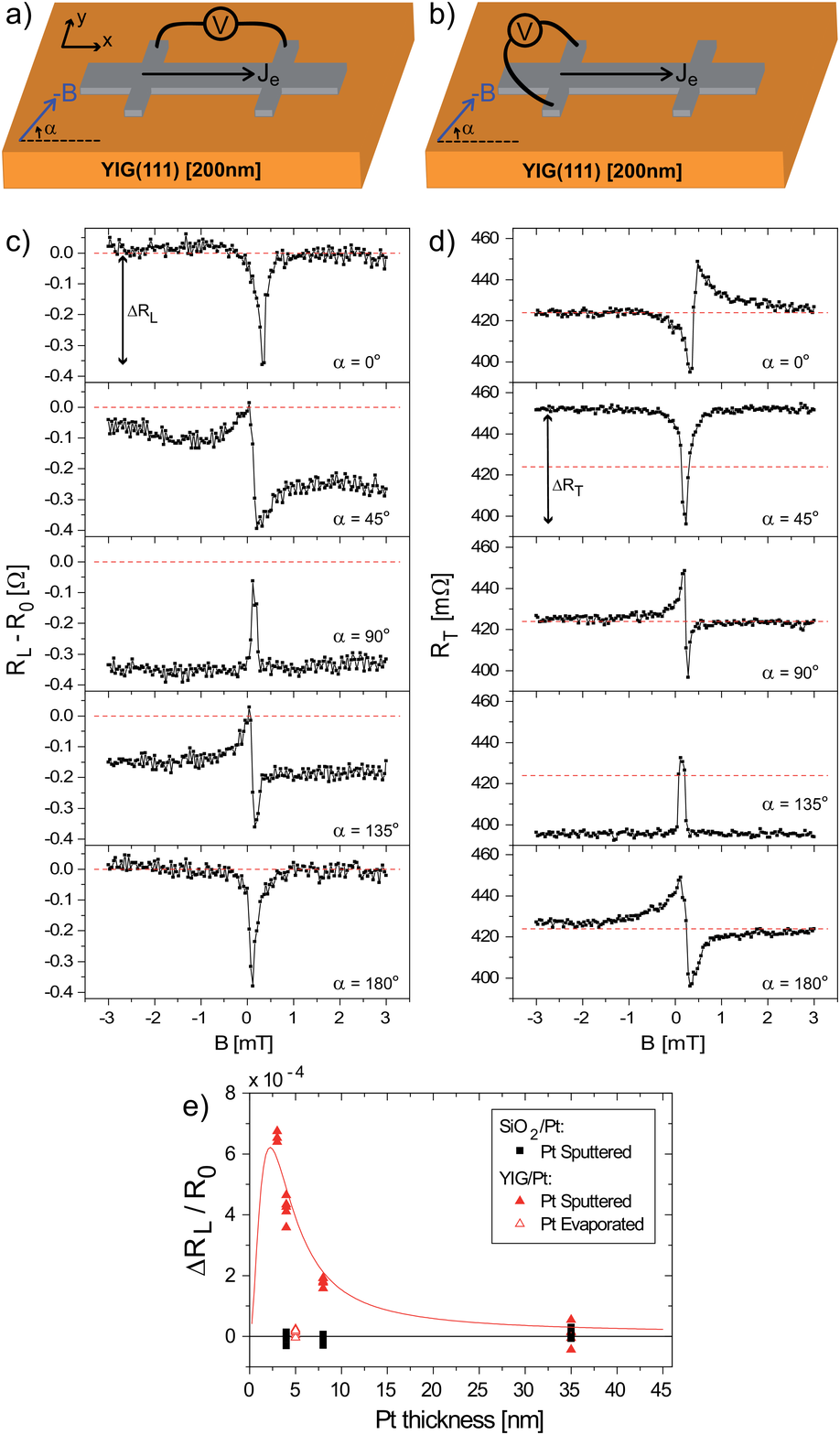}% Here is how to import EPS art
\caption{\label{fig:Fig3} 
Results of the in-plane magnetic field dependence of the resistance of the Pt strip with a thickness of 4nm. Configuration for (a) longitudinal and (b) transverse resistance measurements. (c) and (d) show the measured resistance of the Pt strip while applying an in-plane magnetic field for different angles $\alpha$, for the longitudinal and transverse configuration, respectively. $R_0$ has a magnitude of 1007-1008$\Omega$. (e) Thickness dependence of the measured magnetoresistance for YIG/Pt and SiO$_2$/Pt samples. $\Delta{R_L}$ is defined as the maximum difference in longitudinal resistance ($R_L(\alpha=0^{\circ})-R_L(\alpha=90^{\circ})$) and $R_0$ is $R_L$($\alpha=0^{\circ}$). The solid red line is a theoretical fit.\citep{BauerSMR,BauerTheorySMR}
}
\end{figure}

First, the longitudinal resistance of the Pt strip was measured (using a current $I=100\upmu$A) while sweeping an externally applied in-plane magnetic field. For subsequent measurements the magnetic field was applied for different in-plane angles $\alpha$, as defined in Fig. \ref{fig:Fig3}(a). As the in-plane magnetization of YIG shows isotropic behaviour with a coercive field $B_c$ of only 0.06mT, its magnetization will easily align with the applied in-plane magnetic field. 
It was observed that the measured longitudinal resistance is dependent on the direction of the applied magnetic field, and thus of the magnetization direction of the YIG, as can be seen in Fig. \ref{fig:Fig3}(c) for the YIG/Pt [4nm] sample. For clarity, a background resistance $R_0$ of 1007-1008$\Omega$ was subtracted in the plots (the small change in $R_0$ between different measurements occurred due to thermal drift). A maximum in resistance was observed when the magnetic field was applied parallel to the direction of the charge current $J_e$ ($\alpha=0^{\circ}$). The resistance was minimized for the case where $B$ and $J_e$ were perpendicular ($\alpha=90^{\circ}$). These results are consistent with the SMR as described by Fig. \ref{fig:Fig1} and as observed by Nakayama et al.\citep{BauerSMR}. The measured resistivity for the longitudinal configuration can be formulated as\citep{BauerSMR}
\begin{equation} \label{eq:Rlong}
	\rho_L = \rho_0 - \Delta{\rho} {m_y}^2 \\
\end{equation} 
where $\rho_0$ is a constant resistivity offset, $\Delta{\rho}$ is the magnitude of the resistivity change, which can be calculated from the measurements, giving $\Delta{\rho}=2\times 10^{-10}\Omega$m, and $m_y$ is the component of the magnetization in the $y$-direction.

The same experiments were repeated for the transverse resistance, where the resistance was measured perpendicular to the current path as shown in Fig. \ref{fig:Fig3}(b). Also in this configuration it was found that the measured resistance depends on the direction of the applied in-plane magnetic field, as shown in Fig. \ref{fig:Fig3}(d) for the YIG/Pt [4nm] sample. Here a maximum resistance is observed for $\alpha=45^{\circ}$, and a minimum for $\alpha=135^{\circ}$. The observed SMR resistivity for the transverse configuration can be formulated as\citep{BauerSMR}
\begin{equation} \label{eq:Rtrans}
	\rho_{T} = \Delta{\rho} m_x m_y
\end{equation} 
where $m_x$ is the component of the magnetization in the $x$-direction. From the shown measurements, a ratio $\Delta{R_L}/\Delta{R_T}\approx7$ is found, which is close to the expected ratio of 8 following from equations (\ref{eq:Rlong}) and (\ref{eq:Rtrans}). 

For both the longitudinal and the transverse configuration, there is a peak and/or dip observed around +$B_c$ for all measurements. This can also be explained by the above described SMR. While sweeping the magnetic field (here from negative to positive $B$), the magnetization of the YIG will change direction when passing +$B_c$ (see Fig. \ref{fig:Fig2}(b)). Due to its in-plane shape anisotropy, the magnetization of the YIG will rotate fully in-plane towards $B$. This rotation of $M$ results in a change in measured resistance, passing the maximum and/or minimum possible resistance, which is observed as a peak and/or dip around +$B_c$ (when sweeping the field from positive to negative $B$, a peak/dip will occur at -$B_c$). Similar features were not observed by Huang et al.\citep{ChienPRL2012} and Nakayama et al.\citep{BauerSMR}. They do observe some peaks and dips, but these do not cover the maximum and minimum possible resistances, and thus do not show the full rotation of the magnetization in the plane. The absence of the full peaks and dips can be explained by different magnetization behaviour of their YIG samples, showing higher coercive fields and switching of the magnetization which is probably dominated by non-uniform reversal processes.

The resistance measurements for the in-plane magnetic fields were repeated for all different samples. A summary of these measurements is shown in Fig. \ref{fig:Fig3}(e). Here $\Delta{R_L}$ is defined as the difference between the maximum ($\alpha=0^{\circ}$) and minimum ($\alpha=90^{\circ}$) measured longitudinal resistance and $R_0$ is $R_L(\alpha=0^{\circ})$. The shown thickness dependent measurements are in agreement with data as published by Huang et al.\citep{ChienPRL2012}, though they do not relate their results to SMR. The red line shows a theoretical fit\citep{BauerSMR,BauerTheorySMR} of the SMR signal. The position and width of the peak are mostly determined by the spin relaxation length $\lambda$ of Pt, and the magnitude of the signal by a combination of the spin-Hall angle $\theta_{SH}$ and the spin-mixing conductance $G_{\uparrow\downarrow}$ of the YIG/Pt interface. For the shown fit, $\lambda=1.5$nm, $\theta_{SH}=0.08$, $G_{\uparrow\downarrow}=1.2\times10^{14}\Omega^{-1}$m$^{-2}$ and a thickness dependent electrical conductivity as used in Ref\citep{CastelThicknessPt}, were used. 

When YIG is replaced by SiO$_2$, the SMR signal totally disappears, showing the effect is indeed caused by the magnetic YIG layer. More notable, the e-beam evaporated Pt layer on YIG did show only a very low SMR signal ($\approx10^{-5}$). This suggests that the spin-mixing conductance (which is determined by the interface)\citep{HeinrichInterface} is an important parameter for the occurrence of SMR.

\subsection{Out-of-plane magnetic field dependence}
\begin{figure}
\includegraphics[width=8.5 cm]{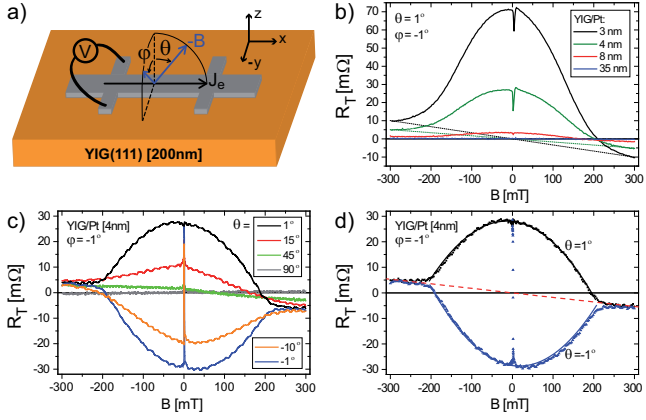}% Here is how to import EPS art
\caption{\label{fig:Fig4} 
Results of the out-of-plane magnetic field dependence of the transverse resistance. (a) Configuration for the transverse resistance measurements. $\phi$ is defined as a rotation from the $z$- towards the -$y$-direction, whereas $\theta$ gives a rotation from the $z$- towards the $x$-direction. (b) Magnetic field dependence of the transverse resistance for different thicknesses of Pt on top YIG, for $\phi=-1^{\circ}$ and $\theta=1^{\circ}$. (c) Dependence of the transverse resistance on $\theta$, fixing $\phi=-1^{\circ}$, pointing out the effect of the direction of the in-plane component of the applied magnetic field on the observed signal. (d) Theoretical fits of the SMR signal for out-of-plane applied fields lower than the saturation field, assuming a linear background resistance, as shown by the dotted red line. For all shown measurements, a constant background resistance of 10-900m$\Omega$ is subtracted.
}
\end{figure}

To further investigate the characteristics of the Pt layer, also the transverse resistance was measured while applying an out-of-plane magnetic field, as shown in Fig. \ref{fig:Fig4}(a). The Pt layers on the Si/SiO$_2$ substrate showed linear behaviour with transverse Hall resistances of 1.3, 0.9 and 0.3 $\pm0.05$m$\Omega$ for Pt thicknesses of 4, 8 and 35nm, respectively, at $B=300$mT. These results, due to the normal Hall effect, are in agreement with the theoretical description $R_{Hall}=R_HB/d$, where $R_H=-0.23\times10^{-10}$m$^3$/C is the Hall coefficient of Pt\citep{Hurd} and $d$ is the Pt thickness.
%The magnitude of the Hall resistance can be theoretically calculated using $R_{Hall}=-B/{ned}=R_H B/d$, where R$_H$ is the Hall coefficient and d is the thickness of the Pt layer. Using this equation, with R$_H$=-0.23 10$^{-10}$ m$^3$/C\citep{Hurd} values of R$_{Hall}$ = 1.15, 0.58 and 0.13 m$\omega$ are found for a Pt thickness of 4, 8 and 35nm respectively. These calculated values are in agreement with the measured transverse voltages, showing that we observe a normal Hall effect, as expected on the SiO$_2$ substrate.

For the YIG/Pt samples, results of the out-of-plane measurements are shown in Fig. \ref{fig:Fig4}(b). At fields lower than the saturation field, a large magnetic field dependence is observed. The magnitude of this dependence decreases with Pt thickness and disappears for the thickest Pt layer of 35nm. 
The occurrence of this magnetic field dependence can be explained by the SMR, using the results of the in-plane measurements as shown in Fig. \ref{fig:Fig3}(d), because for applied fields lower than the saturation field, the magnetization of the YIG will still have an in-plane component. 
To investigate its effect on the transverse resistance measurements, the direction of the in-plane magnetization in the YIG should be known. To achieve this, the external magnetic field was applied with a small intended deviation $\phi$ from the out-of-plane $z$-direction towards the -$y$-direction as defined in Fig. \ref{fig:Fig4}(a). This small deviation results in a small in-plane component of the applied field, which controls the magnetization direction of the YIG. Using this configuration the sign of the signal due to the SMR can be checked according to Fig. \ref{fig:Fig3}(d) by varying the direction of the in-plane component of the applied magnetic field. Fig. \ref{fig:Fig4}(c) shows results applying an external field fixing $\phi=-1^{\circ}$ for various angles $\theta$, where $\theta$ is an additional rotation from the $z$- towards the $x$-direction. According to the theory of the SMR and also comparing the results shown in Fig. \ref{fig:Fig3}(c), a maximum additional resistance due to SMR is expected for an in-plane magnetic field with $\alpha=45^{\circ}$, which is the direction of the in-plane component when applying a magnetic field choosing $\phi=-1^{\circ}$ and $\theta=1^{\circ}$. Similarly, for $\phi=\theta=-1^{\circ}$, the in-plane component of the field will be $\alpha=135^{\circ}$, resulting in a minimum additional resistance. Results as shown in Fig. \ref{fig:Fig4}(c) confirm that the sign and magnitude of the magnetic field dependence are consistent with the SMR observed for in-plane fields. The shape of the curve can be explained by the dependence of the resistance on the direction of $M$, as only the component of ${\sigma}$ parallel to $M$ (${\sigma_M}$) will be reflected. For out-of-plane applied fields, ${\sigma_M}$ is given by ${\sigma_M}={\sigma}\cos{\beta}\cos{\alpha}$, where $\beta$ is the angle by which $M$ is tilted out of the $x/y$-plane. Using the Stoner-Wohlfarth Model,\citep{StonerModel} for an applied field in the z-direction, it was derived that $\beta=\arcsin(b)$, where $b=B/B_s$ and $B_s$ is the saturation field. Assuming that the transverse resistivity change due to SMR scales linearly with the in-plane component of ${\sigma_M}$ ($\sigma_{M,in-plane}=\sigma_M\cos{\beta}$), this gives (for applied fields close towards the $z$-direction and $\phi=\theta=\pm 1$) 
\begin{equation}
	\rho_{T}=\pm \frac{1}{2} \Delta{\rho} (1-b^2)
\end{equation} 

Two fits using this equation are shown in Fig. \ref{fig:Fig4}(d). For both fitted curves, an assumed linear background resistance, as indicated by the dotted red line, is also added. The derived fits are in good agreement with the measured data for applied fields below the saturation field, which confirms the presence of SMR and its dependence on the magnetization direction,\citep{BauerTheorySMR,AlthammerSMR,KleinSMR} also for out-of-plane applied fields.

Also for the out-of-plane measurements a peak and/or a dip is observed at zero applied field. These peaks and dips have the same origin as those observed for the in-plane measurements, which is the rotation of the magnetization in the plane towards the new magnetic field direction.

For applied magnetic fields above the saturation field no in-plane component of M is left, but still a small magnetic field dependence is observed. At $B=300$mT, transverse resistances of 10.1, 5.1, 1.5 and 0.3 $\pm0.05$m$\Omega$ were measured for Pt thicknesses of 3, 4, 8 and 35nm, respectively. So for thin Pt layers, at applied fields above the saturation field, an increased transverse resistance is observed compared to the SiO$_2$/Pt sample. Possible origins of this difference might be related to the imaginary part of the spin-mixing conductance, or to the (spin-) anomalous Hall effect.

\subsection{Comparison of e-beam evaporated and dc sputtered Pt}
\begin{figure}[h]
\includegraphics[width=8.5 cm]{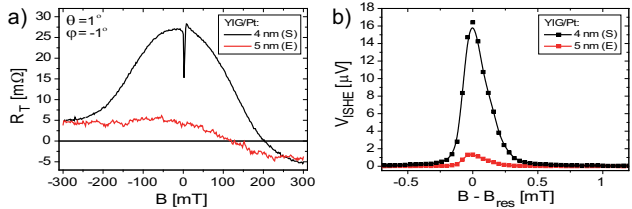}% Here is how to import EPS art
\caption{\label{fig:Fig5} 
Comparison of (a) transverse resistance for an out-of-plane applied magnetic field, and (b) spin pumping/ISHE signal (using an rf-frequency of 1.4GHz with a power of 10mW) for Pt on top of YIG, deposited by e-beam evaporation (E) and dc sputtering (S). 
}
\end{figure}

Additional to the thickness and angular dependence of the SMR signal, also the difference in signal for two deposition techniques, e-beam evaporation and dc sputtering was investigated. It was observed that the e-beam evaporated Pt layer did show very low SMR effects compared to the sputtered layers. To compare, Fig. \ref{fig:Fig5}(a) shows the out-of-plane transverse measurement for both the sputtered [4nm] and evaporated [5nm] Pt layers. The value of the signal at applied fields higher than the saturation field is the same, but the additional signal which is described to SMR is lowered by a factor 7. 

As the evaporated Pt layer showed lower SMR signals compared to the sputtered Pt layers, the effect of using a different deposition technique on the spin pumping/ISHE signal was also investigated. By using a rf-magnetic field, the magnetization of the YIG was brought into resonance. During resonance, a spin current will be pumped into the Pt layer where it will be converted in a charge current by the ISHE. A more detailed description of the used measurement technique can be found in ref.\citep{CastelPRB}.
Fig. \ref{fig:Fig5}(b) shows a measurement of the spin pumping voltage for both e-beam evaporated Pt and dc sputtered Pt on YIG. A rf-frequency and power of 1.4GHz and 10mW, respectively, were used to excite the magnetization precession in the YIG. The same measurement was repeated for different rf-frequencies between 0.6 and 4GHz, all at a power of 10mW (not shown). For all measurements, the spin pumping signal of the evaporated Pt layer was found to be a factor 12 smaller than the signal of the sputtered layer. This change in magnitude of the signal shows the difference of the YIG/Pt interface between both deposition techniques, determining a probable difference in the spin-mixing conductance. As e-beam evaporation is a much softer deposition technique compared to dc sputtering, the spin-mixing conductance at the YIG/Pt interface might be lower in case of evaporation, resulting in less spin pumping.\citep{HeinrichInterface} Also the structure of the Pt layers might be different, resulting in different spin-Hall angles and/or different spin diffusion lengths.

% Summary/Conclusion
\section{Summary}
In summary, the SMR in Pt layers with different thicknesses [3, 4, 8 and 35nm], deposited on top of YIG, was investigated for both in-plane and out-of-plane applied magnetic fields. In-plane magnetic field scans clearly show the presence of SMR for the transverse as well as the longitudinal configuration. Out-of-plane measurements present a magnetic field dependence which can also be assigned to the SMR. The sign and magnitude of the SMR signal are shown to be determined by the magnetization direction of the YIG. Further, thickness dependence experiments show that the SMR signal decreases in magnitude when increasing the Pt thickness. No SMR signals were observed for SiO$_2$/Pt samples. For Pt layers deposited by e-beam evaporation, in stead of dc sputtering, the found SMR signals are decreased by a factor 7. Also spin pumping experiments show reduced signals for e-beam evaporated Pt compared to sputtered Pt. The difference in spin pumping signals and SMR signals show the possible importance of the YIG/Pt interface, and connected to this, the spin-mixing conductance, for this kind of experiments. 

%------------------------------------------------------------------------------------------------------------------------------------------------------------------
%------------------------------------------------------------------------------------------------------------------------------------------------------------------
%------------------------------------------------------------------------------------------------------------------------------------------------------------------

%Acknowledgements
\section*{Acknowledgements}
We would like to acknowledge B. Wolfs, M. de Roosz and J. G. Holstein for technical assistance and prof. dr. ir. G. E. W. Bauer for useful comments regarding the explanation of the measurements. This work is part of the research program (Magnetic Insulator Spintronics) of the Foundation for Fundamental Research on Matter (FOM) and is supported by NanoNextNL, a micro and nanotechnology consortium of the Government of the Netherlands and 130 partners, by NanoLab NL and the Zernike Institute for Advanced Materials.

%\nocite{*}
\bibliography{YIGPt}% Produces the bibliography via BibTeX.

\end{document}